\documentclass[12pt]{iopart}
\usepackage{epsf}
\begin{document}

\article[Chemical equilibration due to heavy Hagedorn states]
{Strangeness in Quark Matter 2004}
{Chemical equilibration due to heavy Hagedorn states}

\author{C.~Greiner\dag\footnote[8]{Invited speaker.},\ 
P.~Koch-Steinheimer\dag\ddag,\
F.~M.~Liu\dag\S,\
I.~A.~Shovkovy\dag\P\footnote[7]{On leave of absence from
       Bogolyubov Institute for Theoretical Physics,
       03143, Kiev, Ukraine.}
and H.~St\"ocker\dag\P}

\address{\dag\ Institut f\"ur Theoretische Physik,
Johann Wolfgang Goethe-Universit\"{a}t,
D-60054 Frankurt am Main, Germany}

\address{\ddag\ Bitfabrik, D-60314 Frankurt am Main, Germany}

\address{\S\ Institute of Particle Physics, 
Central China Normal University, 430079 Wuhan, China}

\address{\P\ Frankfurt Institute for Advanced Studies (FIAS),
Johann Wolfgang Goethe-Universit\"{a}t,
D-60054 Frankurt am Main, Germany}

\date{\today}

\begin{abstract}
A scenario of heavy resonances, called massive Hagedorn states,
is proposed which exhibits a fast ($t\approx 1$~fm/c) chemical 
equilibration of (strange) baryons and anti-baryons at the QCD 
critical temperature $T_c$. For relativistic heavy ion collisions
this scenario predicts that hadronization is followed by a brief 
expansion phase during which the equilibration rate is higher 
than the expansion rate, so that baryons and antibaryons reach 
chemical equilibrium before chemical freeze-out occurs.
\end{abstract}



\submitto{\JPG}


\section{Introduction}

The enhancement of (multi-)strange (anti-)baryons has been predicted 
as a potential proof for the existence of the quark gluon plasma (QGP) 
in relativistic heavy ion collisions \cite{KMR86}. Hadro-chemically 
saturated multiplicities of (anti-)hyperons have been experimentally 
discovered in central collisions in Pb+Pb experiments at CERN-SPS and 
in Au+Au experiments at Brookhaven, see Ref.~\cite{BSR}.

At SPS energies, the strong increase in the yield of antiprotons 
\cite{RS00} and antihyperons \cite{CAR1,CAR2} has been explained
by a ``clustering'' of mesons of the following type:
\begin{equation}
\label{mesfuse}
n_1\pi + n_2 K \, \leftrightarrow \, \bar{Y}+p \, .
\end{equation}
This channel is very efficient to chemically equilibrate (strange) 
antibaryons in a baryon-dense fireball as is expected at the SPS 
energies. The equilibration time scale is inversely proportional 
to the density of baryons and to the annihilation cross-section, i.e.,
$(\Gamma _{\bar{Y}})^{(-1)} \sim 1/ (\sigma _{B \bar{Y}} v _{B\bar{Y}} 
n_B )$. Assuming that various cross-sections are related as given 
roughly by the additive quark model, $\sigma _{B \bar{Y}\rightarrow 
n  \pi + n_Y K} \approx \sigma _{N \bar{p}\rightarrow n  \pi } $,
(for a theoretical model, see Ref.~\cite{IGOR}) and that the density 
is about 1 to 2 times larger than the normal nuclear matter density 
$\rho_0 $, one finds equilibration times for antibaryons $t_{\rm eq }
\approx 1$--$3$~fm/c \cite{CAR1}. This leads to microscopically calculated 
antihyperon yields which are consistent with the fitted chemical 
freeze-out parameters \cite{CAR2}.

The multi-meson reactions at RHIC cannot account for an efficient, 
direct baryon-antibaryon pair production within the standard hadron 
resonance gas model. The observed ``chemically saturated'' (strange) 
(anti-)baryon yields \cite{RS00,CAR1,IGOR} cannot be obtained from
hadron cascades. Antibaryons are much more abundant at RHIC than at 
SPS, and equilibration times are $\sim 10$~fm/c at a temperature of 
about 170 MeV \cite{CAR1,IGOR,PASI}. Nevertheless, the RHIC hadron 
multiplicities suggest that chemical equilibrium is reached at about 
$170$~MeV~$\approx T_c$. Thus, it has been suggested that various 
hadron species are ``born in equilibrium'' \cite{Stock}. 

Here, a scenario of dynamical equilibration is developed, overcoming 
some shortcomings (see Sec.~\ref{shrt}) of Ref.~\cite{BSW} (hereafter
referred to as BSW). According to BSW, the reactions in Eq.~(\ref{mesfuse})
alone lead to chemical equilibration at RHIC. Sec.~\ref{new-sce}
presents a sufficiently fast ($\approx 1$~fm/c) chemical equilibration 
mechanism for both of (strange) baryons and anti-baryons. It relies on 
abundant (at $T_c$) massive resonances, called Hagedorn states (HS), 
which are short lived. (Strange) baryon and antibaryon production 
proceeds along the reaction
\begin{equation}
\label{HSdec}
\left( n_1\pi + n_2 K + n_3 \bar{K} \, \leftrightarrow \right) \, HS \, 
\leftrightarrow \bar{B}+B+X \, .
\end{equation}
Here $X$ represents all possible multi-hadron states. Such reactions 
are an important generalization of Eq.~(\ref{mesfuse}). The equilibration 
time can be estimated by the branching ratios of HS decays into 
$B+\bar{B}+X$ using a microcanonical statistical model. We then address 
the production of strange (anti-)baryons, especially the rare $\Omega$ 
(sss) state. The role of HS for chemical equilibration at nonzero 
net baryon density (e.g., at SPS and AGS energies) is briefly 
discussed.

\section{Overpopulated hadron densities}
\label{shrt}

Typical equilibrium baryon/antibaryon densities in a net baryon-free 
hadronic system at $T\approx 170-180$~MeV are $n_B^{eq} = n_{\bar{B}}^{eq}
\approx 0.04~\mbox{fm}^{-3}$ (neglecting eigenvolume effects). With 
annihilation cross sections $\langle\sigma v \rangle\approx 30$~mb, 
the equilibration time due to reactions as in Eq.~(\ref{mesfuse}) is 
$t_{\rm eq}\approx 10$~fm/c. This cannot explain the apparent chemical 
equilibration of baryon and antibaryons \cite{CAR1,IGOR,PASI}. It was 
suggested in Ref.~\cite{BSW} that such multi-meson collisions can still 
lead to quick chemical equilibration, in close vicinity to the phase 
transition: by comparing a hadron gas model equation of state with the 
lattice results, it was speculated that there exists a state of 
extra-large particle density around $T_c$, which is effectively 
overpopulated with pions and kaons. This overpopulation then could drive 
the baryon and antibaryon pair production on a very short time scale. 
In essence, the total production of a specific type of baryon (e.g., 
the $\Omega$ as the most exotic one) is attributed \cite{BSW} to the 
gain term in the following master equation \cite{CAR1,CAR2,IGOR,PASI}:
\begin{equation}
\hspace*{-22mm}
\frac{d}{dt}n_{\Omega} \, =\,  -\mbox{``loss''}+\mbox{``gain''}
\equiv -\langle \sigma_{\Omega\bar{B}} v_{\Omega\bar{B}}\rangle 
\times 
\left(n_{\bar{B}}n_{\Omega}-\sum_{n_1,n_2}\hat{M}_{(n_1,n_2)}
(n_{\pi})^{n_1}(n_{K})^{n_2}\right) \, .
\label{bal-1}
\end{equation}
Here the mass-action factor reads
$
\hat{M}_{(n_1,n_2)} ={n^{(eq)}_{\bar{B}}n^{(eq)}_{\Omega}}/
{(n_{\pi}^{(eq)})^{n_1}(n_{K}^{(eq)})^{n_2}}.
$
Notice that, in accord with Eq.~(\ref{bal-1}), the density of $\Omega$s
tends to approach $n^{(eq)}_{\Omega}$ during chemical equilibration. 
If the equilibrium is maintained, i.e., when  the ``loss'' and ``gain''
reactions have equal rates, saturation does not change. In this case, the 
production (as well as the annihilation) rate are given by
$
\Gamma_\Omega \simeq 
\langle \sigma_{\Omega\bar{B}} v_{\Omega\bar{B}}\rangle
n_{\bar{B}},
$
and the corresponding chemical equilibration time is $\tau^0_\Omega 
\simeq 1/\Gamma_\Omega$. This gives $\tau^0_\Omega \geq 10~\mbox{fm/c}$. 
 
The master equation (\ref{bal-1}) is valid even if the particle yields
are initially not in full chemical equilibrium. It then dictates how 
the population changes with time. The BSW idea can now be formulated
as follows: the overpopulation of meson states by a factor $\alpha$,
$n_{\pi,K} \rightarrow \alpha \cdot n_{\pi,K}^{eq}$, as compared to 
the standard equilibrium value in the hadron resonance gas model at
$T_c$ should result in rates rescaled by $\alpha^5$, i.e.,
$ \Gamma _{\Omega }^{prod}\, = \, \alpha^5 \cdot \Gamma^{eq}_{2\pi 3K 
\rightarrow  \Omega  \bar{B}}$. If $\alpha =2$, the production rate 
for $\Omega$s (as well as for other baryons and antibaryons) increases 
by a factor 32! One concludes, therefore, that baryons and antibaryons 
are readily produced by multi-meson reactions in an overpopulated bath 
of pions and kaons, and the equilibration should happen on timescales 
$t_{\rm eq} \stackrel{<}{\sim } 1$~fm/c \cite{BSW}. The rapid ($\sim 
T^{60}$) fall-off \cite{BSW} of the multi-particle reaction rates 
with temperature suggests that the ``chemical freeze-out temperature'' 
is also a measure of the phase transition temperature.

However, there is a serious consequence of this argument: literally, 
it predicts that baryons and antibaryons are tremendously overpopulated 
as compared to their equilibrium values. Once produced, there is no way 
to get rid of them any more in the later stages, because the annihilation 
reactions are then suppressed dynamically. This is not seen in 
experiment. Inspect once again the master equation (\ref{bal-1}): 
its (quasi-) stationary fixed point is reached when the 
expression in parentheses vanishes, i.e., when the annihilation 
rate is equal to the production rate. This corresponds to (anti-)baryon 
densities, rescaled by a factor $\beta$, i.e., $n_{\Omega, \bar{B}} 
\rightarrow \beta \cdot n^{eq}_{\Omega, \bar{B}}$, where $\beta = 
(\alpha )^{5/2} $. This may slightly change when other multi-meson 
channels are taken into account. For $\alpha = 2$, $\beta=5.6$, which 
predicts way too many (anti-)baryons in the system. In fact, the state 
with overpopulated (anti-)baryons is reached on a timescale of 
$\tau = \tau^0_\Omega/ \beta $. This is short enough to compensate 
the rapid fireball expansion. However, once a large number of 
(anti-)baryons is produced, it is difficult to get rid of them quickly 
enough in order to reach standard hadron equilibrium values before the 
chemical freeze-out. The fraction of (anti-)baryons which annihilates
is too small, because the corresponding annihilation rates are then 
not sufficiently large in an ordinary hadron resonance gas.

\section{Fast equilibration due to Hagedorn States}
\label{new-sce}

We propose to circumvent this drawbacks of the BSW scenario by postulating
that the needed additional degrees of freedom close to $T_c$ are not
light mesons (``pions'' and ``kaons''), as proposed in Ref.~\cite{BSW}, 
but that they consist of heavy mesonic and baryonic resonances, called 
Hagedorn states. The conjectured unstable Hagedorn states can produce 
sufficiently many baryon-antibaryon pairs, on a sufficiently short time 
scale. HS efficiently and effectively can account for various 
multi-particle collisions (``interactions'') in a consistent way: 
\begin{eqnarray}
HS & \leftrightarrow & 
n_1 \cdot \pi + n_2 \cdot K + n_3\cdot \bar{K}\, , \\
HS & \leftrightarrow &
B  +  
\bar{B}\, , \\
HS &\leftrightarrow &
B  + 
\bar{B}  
+ \bar{n}_1 \cdot \pi + \bar{n}_2 \cdot K + \bar{n}_3\cdot \bar{K}
\, \equiv \, B  + 
\bar{B} + X \, \, .   
\label{HSdecay}
\end{eqnarray}
This is akin to the Hagedorn's original idea \cite{Hag}, that the 
strong interactions of low mass hadrons can be attributed to an 
exponentially increasing mass spectrum as $T\to T_c$. This old idea 
has severe consequences for the time-scales of chemical equilibration 
--- from the phase space arguments, one intuitively expects that 
multi-particle decays in (\ref{HSdecay}) dominate the $B \bar{B}$ 
production.

Some comments to the Hagedorn picture are in order. Close to the 
crossover (critical) temperature, $T_c$, the conventional description 
of QCD in terms of the known set of PDG hadron degrees of freedom can 
indeed not be sufficient (PDG: all known resonances from the particle 
data group booklet). In fact, one should insist that this set is
not sufficient in order to understand the lattice results on the 
equation of state of QCD \cite{BSW,RGM,BB}. The PDG-hadron resonance 
gas model falls short when estimating the pressure of lattice QCD as 
the crossover temperature is approached. This is not surprising, if 
one recalls the role of the highly lying massive states in the Hagedorn 
model. In the absence of a clear phase transition, it is easy to 
imagine that (additional, unknown) hadronic degrees of freedom can 
be used in the description of hot matter, in fact, even for temperatures 
above the crossover temperature \cite{Schramm}. Of course, the further 
one goes away from ``$T_c$'', the less convenient such descriptions 
should become. Physically, this could be due to the increasing width 
of resonances \cite{BB}. In the standard Hagedorn model, the density 
of states increases exponentially with their mass. Eventually, this 
leads to a divergence of the pressure, as the value of the temperature 
approaches the Hagedorn temperature from below. Is the Hagedorn model
ruled out, therefore? We believe this is not necessarily the case.
As pointed out in Ref.~\cite{BB,GOR}, the model can be modified in a 
natural way that cures the problem with the pressure. In addition, 
quite recently the philosophy of Hagedorn behaviour has also become 
popular with regard to the equation of state of large N gauge theories 
\cite{Sen} and Super-Yang-Mills theories \cite{SYM}.

\subsection{Estimate of the baryon/antibaryon production}

The number density of HS states in the vicinity of the critical 
temperature is estimated as in Ref.~\cite{BSW}: additional degrees of 
freedom to the PDG hadron resonance gas are needed in order to understand 
the energy density of a thermal system around $T_c$, obtained in lattice 
QCD calculations. The additional energy density of order $\Delta 
\epsilon_c  \approx 0.3-0.5 $~GeV/fm$^{3}$ is needed. Let us identify 
this with the energy density of the HS, $\Delta \epsilon_c \equiv 
\epsilon _{HS}$. Than, a typical HS mass is $ M_{HS}\approx 3$--$6$~GeV. 
The total number density of these HS is 
\begin{equation} 
n_{HS } \approx  \frac{ \Delta \epsilon _c }{ \langle M_{HS}\rangle}
\approx 0.05 -0.15 \, \mbox{fm}^{-3} \, \, .
\end{equation}
We now proceed to estimate the baryon-antibaryon production rate
due to $HS  \rightarrow B \bar{B}  + X$, where $X$ stands for any 
possible number of additional hadrons. HS must be highly unstable: 
the phase space for multi-particle decays becomes immense with 
increasing mass. Extrapolating the width from the known meson 
resonances at 2 GeV (widths of $0.3$--$0.5$~GeV) linearly (as 
suggested by the string model, e.g., in the last paper of 
Ref.~\cite{Sen}), we find that the total width of such high mass 
mesonic states is $\Gamma ^{tot}_{HS }\stackrel{>}{\sim} 0.5-1$~GeV. 
In the next subsection, we estimate the average baryon number 
$\langle B \rangle$ per unit decay of a HS within a microcanonical 
approach. Here we quote only the result: $\langle B \rangle \approx 
0.2-0.4$. Hence, the relative decay width is given by 
$\Gamma_{B \bar{B} X} \approx \langle B \rangle \cdot \Gamma^{tot}_{HS} 
\approx 100$--$300$~MeV. Take the lower value in the following. Then, 
the production rate for baryon-antibaryon pairs is estimated
\begin{equation}
\frac{dN_{ B \bar{B} }}{d^4x} = 
\Gamma^{prod}_{ B \bar{B} }  =  n_{HS } \cdot
\Gamma _{HS  \rightarrow B \bar{B} + X} \approx 0.05 \, \mbox{fm}^{-4} \, .
\end{equation}   
With $n_{\bar{B}}^{eq} \approx n_B^{eq} \approx 0.04 \, \mbox{fm}^{-3}$ 
at chemical freeze-out, at RHIC, for all baryons and anti-baryons, 
their chemical equilibration rate is given via 
$\Gamma^{prod}_{B \bar{B}}/{n_{ \bar{B} }^{eq}}\approx 1.25 \, 
\mbox{fm}^{-1} $: the chemical equilibration time is very 
small, $\tau^{chem}_{B \bar{B}} \approx 0.8$~fm/c. 

A large fraction of all possible reactions (\ref{HSdecay}) also generates 
chemical equilibration among the ``lighter'' hadrons. For example, these
distinct reaction channels easily alter the number of pions (or kaons) 
in the system (e.g., $7 \pi \leftrightarrow 4 \pi $), so that in turn 
the light degrees of freedom become fully equilibrated chemically. 

\begin{figure}[htp]
\centerline{\epsfysize=10cm \epsfbox{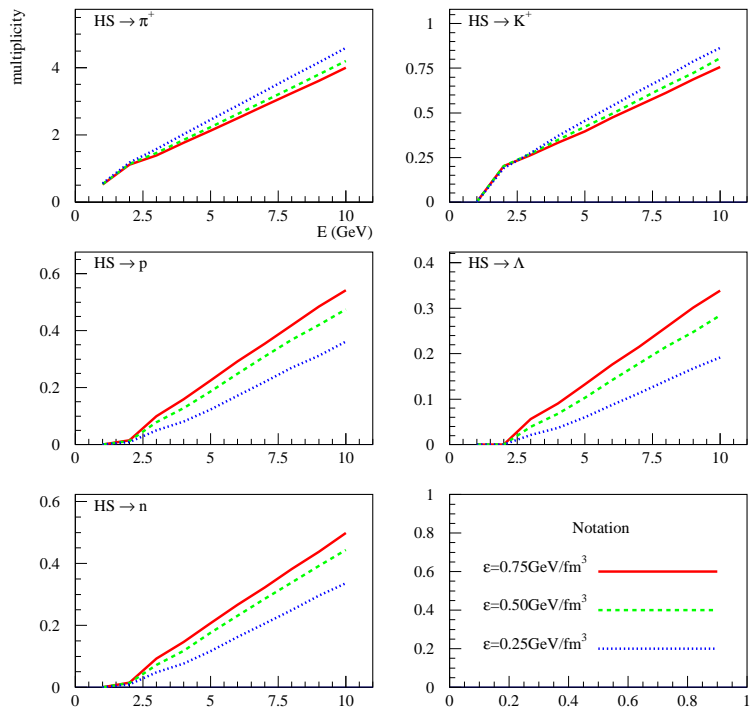}}
\caption{Multiplicity of individual hadrons coming per unit decay
of a mesonic, nonstrange HS as a function of the mass $E=M_{HS}$.}
\label{Fuming}
\end{figure}

The population of HS should die out very rapidly when the temperature 
is decreased within a narrow interval around the critical point. Thus, 
our scenario predicts that the chemical decoupling should happen 
naturally very close to $T_c$, similarly as it was anticipated in the 
BSW scenario \cite{BSW}. In contrast to the BSW mechanism, however, the 
approach proposed here does {\it not} lead to the BSW-oversaturation
of baryons and antibaryons from decaying HS. Indeed, one might worry 
that such decays could produce too many $B\bar{B}$ pairs during the 
expansion and cooling of fireball. This is no problem as soon as 
the cooling proceeds smoothly, the densities of baryons and 
antibaryons stay close to their equilibrium values. However, in the 
extreme case of ultrafast cooling and decoupling, the number of 
additionally produced pairs (via the decay of all HS states) can be 
estimated to be $\delta n_{B \bar{B} } \stackrel{<}{\sim}
(\Gamma _{B \bar{B} + X}/ \Gamma ^{tot}_{HS }) \cdot n_{HS} \approx 
0.2 \, n_{HS}$. Thus, relative overpopulation is $\delta 
n_{B \bar{B}} / n^{eq}_{B} \stackrel{<}{\sim} 0.2$--$0.4$. 
As this is the most extreme case, the actual overpopulation will 
certainly be much smaller.

\subsection{Microcanonical decay of a Hagedorn state}

We now provide an estimate for the individual branching ratios 
of the decay of a HS state into baryon-antibaryon pairs. The 
$B \bar{B}$ annihilation reactions at LEAR have been understood 
well within a statistical description \cite{Van89}. Without any 
further information, we assume that the HS decay also with a 
statistical, microcanonical branching.

\begin{figure}[htp]
\centerline{\epsfysize=6cm \epsfbox{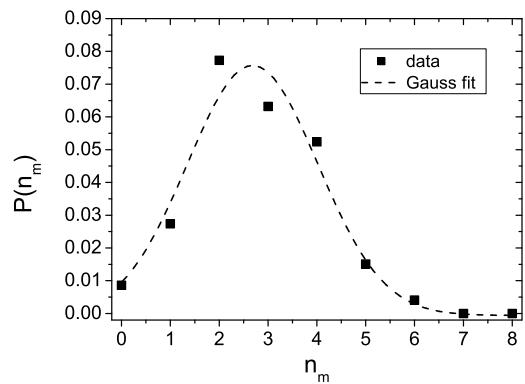}}
\caption{Branching probability distribution of additionally
produced mesons $HS\rightarrow B+\bar{B} + n_m\cdot M$ as a 
function of the meson number per unit decay of a $M_{HS}=4$ GeV 
Hagedorn state.}
\label{Dist}
\end{figure}

The microcanonical description of hadron reactions (e.g., $pp$ or 
$p \bar{p}$) developed in Ref.~\cite{Liu} is used here. A state is 
described solely by its mass $m\equiv M_{HS}$, by its (reaction) 
volume $V\equiv V_{HS}= M_{HS} / \epsilon $, where $\epsilon$ 
denotes the mean energy density of a HS state, and by quantum 
numbers such as the net baryon number and the net strangeness.

In Fig.~\ref{Fuming} we depict the various numbers of hadrons $h$ 
in the distinct decay channels $HS \rightarrow h + X$ per unit decay. 
The HS states are purely mesonic, with no net strangeness. The energy 
density $\epsilon$ of the HS states is varied in the range 
$0.25$--$0.75 $ GeV/fm$^3$. The mass of the HS is varied continuously 
from $M_{HS} =0$~GeV to $10$~GeV. For example, in a single decay of 
a state with mass $m_{HS}=4$ GeV, there are about $2$ $\pi^{+}$, 
$0.075$--$0.15$ protons and neutrons each, $0.35$ $K^{+}$ and 
$0.025$--$0.075$ $\Lambda$'s produced. For the mean baryon number, 
one finds $\langle B \rangle=0.2$--$0.4$. This was the input value 
used in the estimate of the partial decay width in the previous 
subsection.

Fig.~\ref{Dist} shows the branching distribution $p(n_m)$ of the 
associated mesons as a function of the meson number per unit decay 
of a $M_{HS}=4$~GeV Hagedorn state. They are produced in 
$HS\rightarrow B +\bar{B} + n_m\cdot M$. The HS energy density is 
$\epsilon=0.5$~GeV/fm$^3$. The sum of the branching probabilities 
adds up to $\langle B\rangle  \approx 0.3$, as noted above. On average 
3 (stable) mesons accompany a HS decay into a baryon-antibaryon pair. 
Moreover, the direct decay, $HS \rightarrow B+\bar{B}$, is {\em not} 
likely.

\subsection{Strange baryons and antibaryons }

The $\Lambda $ particle has a sizeable production probability in the 
decay of nonstrange mesonic HS. One can estimate similarly that the 
$\Lambda $ and the $\bar{\Lambda }$ can be quickly populated and 
reach their chemical equilibrium value by the interplay between 
the decay and the production of the HS. What about the most exotic 
baryonic state, the $\Omega$, with three units of strangeness?

First, one might think that the $\Omega$ can also be produced 
directly from nonstrange HS, i.e., via $HS _{{\footnotesize 
\mbox{ nonstrange} }} \, \leftrightarrow \Omega  + \bar{B} + X $. 
For a $M_{HS}=4$ GeV HS, the average branching into $\Omega$s per
unit decay is calculated to be $\langle\Omega\rangle \approx 2\cdot 
10^{-4} $, so $\Gamma^{prod}_{\Omega } \approx 0.1-0.2$ MeV.
At $T=170$~MeV, the density of $\Omega $s in a PDG hadron gas
is $n_{\Omega }^{eq} \approx 4\cdot 10^{-4}$ fm$^{-3}$.
The chemical equilibration rate is
\begin{equation}
\Gamma^{chem}_{\Omega } =
\Gamma^{prod}_{\Omega }\,  \frac{n_{HS }}{n_{\Omega }^{eq}}
\approx 25-50 \, \mbox{MeV} \, 
\Rightarrow \, \tau_{\Omega } \approx 4-8 \, \mbox{fm/c} \, .
\end{equation}
This timescale is too long to (fully) explain a chemically
saturated abundance.

As a next step, consider {\em (multi-)strange } mesonic HS, for which 
the reaction $HS  (sss \bar{q} \bar{q} \bar{q} )\, \leftrightarrow 
\Omega + \bar{B}  + X $ might be sufficient. It turns out that 
$\langle \Omega \rangle \approx 0.05 $ for a $m_{HS}=4$ GeV and thus 
$\Gamma^{prod}_{\Omega } \approx 25-50 $
MeV. For the chemical equilibration rate one has
\begin{equation}
\Gamma^{chem}_{\Omega } =
\Gamma^{prod}_{ \Omega  }\,  
\frac{n_{HS }(sss\bar{q} \bar{q} \bar{q} )}{n_{\Omega }^{eq}}
\approx 50-100 \, \mbox MeV \quad \Rightarrow  \tau_{\Omega } \approx 2-4 \,
\mbox{fm/c } \, .
\end{equation}
Here we have assumed that $n_{HS}(sss\bar{q} \bar{q} \bar{q}) /n_{\Omega } 
\approx n_{HS\!{{\scriptsize \mbox{ nonstrange} }}}/n_{B } \approx 2.5 $.
Altogether, strange Hagedorn states can explain the population of $\Omega$.

\subsection{Importance of baryonic HS: from the Hagedorn temperature
to the Hagedorn line?}

So far, we have only considered the role of mesonic HS. Now we 
discuss the role of baryonic HS. Until now, we have concentrated 
on the mid-rapidity region at RHIC, where $\mu _B \approx 0 $. 
Here, the fraction of baryonic to mesonic states can be estimated 
by comparison with that for normal hadrons:
$n_{BHS }/n_{MHS} \approx n_{B}/n_{M} \approx 0.04 / 0.3 = 0.13$.
One thus would conclude that these states do not carry a major fraction 
of the energy density. This can change, though, at finite $\mu_B$ at AGS 
or SPS regime, where the baryonic HS achieve an extra enhancement factor 
$e^{\mu_B/T }$. Such states could show up along  a narrow $\mu-T$ band 
below a critical ``Hagedorn'' line $T_H(\mu_B )$ \cite{HagRaf}. This 
conjecture requires further development that is beyond the scope of 
this paper.

Baryonic HS can also help to produce strange baryons. Consider the 
$\Omega $ and any $\Omega$-like HS, respectively. The latter can decay 
and can be produced by the following two reactions:
\begin{eqnarray}
(a)  \quad \Omega HS & 
\leftrightarrow & B\left( \neq \Omega \right)  + X  
\label{OHSa} \\
(b)  \quad \Omega HS & 
\leftrightarrow & \Omega   + X  \label{OHSb}
\end{eqnarray}
Here $ B \neq \Omega $ denotes baryon states with net strangeness
less than three. The HS with higher net strangeness are populated 
by multi-hadronic fusion of baryons (resonances) with lesser strangeness 
and with other mesons, but with at least one kaon. On the other hand, 
the $\Omega$s then stem from the decay of these $\Omega HS$. If such 
a scenario is to work, the $\Omega HS$ have to be produced sufficiently 
fastly. We now  estimate the chemical equilibration times, both for 
saturating the $\Omega HS$ and for saturating the $\Omega$s. To do this 
one needs the relative branching probability separately for the decays 
happening via (a) and via (b). By employing the microcanonical model, 
one obtains for a $M_{\Omega HS}=4$ GeV Hagedorn state $p_{(a)}=0.9$ 
and  $p_{(b)}=0.1$. The $\Omega HS$ predominantly decays into baryons 
with lesser strangeness. The chemical equilibration time for the 
$\Omega HS$ can be written
\begin{equation}
\label{tau1}
 \left( \tau^{chem}_{ \Omega  HS } 
\right) ^{-1} = \Gamma _{ \Omega  HS   
\leftrightarrow  B  + X }  \approx p_{(a)} \cdot \Gamma ^{tot}= 
0.4-0.8 \, GeV \, .
\end{equation} 
Hence, such states do equilibrate very fast.
For the $\Omega $ we have
\begin{equation}
\label{tau2}
\left( \tau^{chem}_{ \Omega } 
\right) ^{-1} = \Gamma _{ \Omega HS  
\leftrightarrow   \Omega   + X } \, 
\frac{n_{\Omega HS  }}{n_{ \Omega  }^{eq}}  =  
p_{(b)} \cdot \Gamma ^{tot} \, 
\frac{n_{\Omega HS  }}{n_{ \Omega  }^{eq}} \, .
\end{equation}
The ratio of $\Omega $-like resonances to $\Omega$ in the vicinity 
of $T_c$ is probably $O(1)$. We estimate $n_B/(n_p + n_n)\approx 4$ 
at T=170 MeV. For $n_{\Omega HS  }/n_{ \Omega  }^{eq}=1$, one has  
$\tau_{ \Omega } \sim 2-4$ fm/c. Hence, exotic $\Omega HS$ can also 
be a key for explaining the chemical saturation of the $\Omega $.
This would also be true, in particular, at SPS energies.
Such a mechanism can work at finite $\mu_B $, as the
ratio $n_{\Omega HS  }/n_{ \Omega  }^{eq} $ does not depend on 
$\mu_B$ in Boltzmann approximation.

\section{Summary and Conclusions}

We have elaborated the special role of Hagedorn states for chemical 
equilibration of (strange) baryons and antibaryons. The chain of 
reactions (\ref{HSdec}) or (\ref{HSdecay}) catalyzes rapid 
equilibration of antibaryons and baryons in the vicinity of 
the deconfinement transition. The production and the decays of HS 
are governed by detailed balance, where both the continuous repopulation 
of HS as well as the annihilation of baryon-antibaryon pairs in
the back reactions drives their chemical saturation.
Three assumptions are necessary:
(i)
$\Delta \epsilon _{HS } \approx 0.3-0.5$~GeV/fm$^3 $ at $T\approx T_c$;
(ii) 
$\Gamma^{tot}_{HS } \stackrel{>}{\sim} 0.5-1$~GeV; 
(iii) 
a microcanonical, statistical estimate of individual branching ratios.

The Hagedorn states are additional degrees of freedom which represent 
complicated many-particle (hadronic or partonic) plasma correlations 
in hot dense matter. Close to the critical temperature this plasma is 
a strongly interacting phase, which contains such states. A clear cut 
proof, however, within the lattice QCD is not available at present
\cite{footnote}.

\section*{Acknowledgements}
This work was supported by Gesellschaft f\"{u}r Schwerionenforschung 
(GSI) and by Bundesministerium f\"{u}r Bildung und Forschung (BMBF).
The work of F.M.L. was supported in part by the Alexander von 
Humboldt-Foundation.


\section*{References}

\begin{thebibliography}{99}

\bibitem{KMR86} 
P.~Koch, B.~M\"uller and J.~Rafelski,
Phys.\ Rep.\ {\bf 142} (1986) 167.

\bibitem{BSR} 
P.~Braun-Munzinger, K.~Redlich and J.~Stachel,
nucl-th/0304013, in {\it Quark gluon plasma III}, 
edited by R.C.~Hwa and Xin-Nian Wang (World Scientific,
Singapore, 2004) pp. 491-599.  

\bibitem{RS00} 
R.~Rapp and E.~Shuryak, 
Phys.\ Rev.\ Lett.\ {\bf 86} (2001) 2980.

\bibitem{CAR1}
C.~Greiner and S.~Leupold,
J.\ Phys.\ G {\bf 27} (2001) L95.

\bibitem{CAR2} C.~Greiner,
AIP Conf.\ Proc.\  {\bf 644} (2003) 337.

\bibitem{IGOR}
J.~Kapusta and I.~A.~Shovkovy,
Phys.\ Rev.\ C {\bf 68} (2003) 014901.

\bibitem{PASI}
P.~Huovinen and J.~Kapusta, 
Phys.\ Rev.\ C {\bf 69} (2004) 014902.

\bibitem{Stock}
R.~Stock,
Phys.\ Lett.\ B {\bf 456}, 277 (1999).

\bibitem{BSW}
P.~Braun-Munzinger, J.~Stachel and C.~Wetterich,
Phys.\ Lett.\ {\bf B596} (2004) 61.

\bibitem{Hag} R.~Hagedorn, Nuovo Cim. Suppl. {\bf 3} (1965) 147;\\
R.~Hagedorn and J.~Ranft, Nuovo Cim. Suppl. {\bf 6} (1968) 169.

\bibitem{RGM}
F.~Karsch, K.~Redlich and A.~Tawfik,
Eur.\ Phys.\ J.\ C {\bf 29} (2003) 549;
Phys.\ Lett.\ B {\bf 571} (2003) 67.

\bibitem{BB}
D.~B.~Blaschke and K.~A.~Bugaev,
nucl-th/0311021.

\bibitem{Schramm} S.~Schramm and M.~C.~Chu,
Phys.\ Rev.\ D {\bf 48} (1993) 2279.

\bibitem{GOR}
M.I.~Gorenstein, W.~Greiner and S.N.~Yang,
J.\ Phys.\ G {\bf 24} (1998) 725.

\bibitem{Sen} F.~Lizzi and I.~Senda, Phys.\ Lett.\ B{\bf 244} (1990) 27;
Nucl.\ Phys.\ B{\bf 359} (1991) 441; \\
I.~Senda, Phys.\ Lett.\ B{\bf 263} (1991) 270.

\bibitem{SYM} B.~Sundborg,
Nucl.\ Phys.\ B {\bf 573} (2000) 349;
O.~Aharony, J.~Marsano, S.~Minwalla, K.~Papadodimas and M.~Van Raamsdonk,
hep-th/0310285.

\bibitem{Van89} J.~Cugnon and J.~Vandermeulen,
Phys.\ Rev.\ C {\bf 39} (1989) 181.

\bibitem{Liu}
F.~M.~Liu, K.~Werner and J.~Aichelin,
Phys.\ Rev.\ C {\bf 68} (2003) 024905;\\
F.~M.~Liu, K.~Werner, J.~Aichelin, M.~Bleicher and H.~St\"ocker,
J.\ Phys.\ G {\bf 30} (2004) S589;\\
F.~M.~Liu, J.~Aichelin, K.~Werner and M.~Bleicher,
Phys.\ Rev.\ C {\bf 69} (2004) 054002.

\bibitem{HagRaf}
R.~Hagedorn and J.~Rafelski,
in {\it Statistical Mechanics of Quarks and Hadrons}, 
edited by H.~Satz (North-Holland, Amsterdam, 1981) 
pp. 237-251.

\bibitem{footnote} Studies along the lines of Ref.~\cite{Schramm} by 
S.~Schramm are in progress.

\end{thebibliography}
\end{document}